\documentclass[11pt]{article}
\usepackage{amssymb,amsmath}
\usepackage{graphicx}
\usepackage{dcolumn}
\usepackage{bm}
\usepackage{fullpage}
\usepackage{color}
\begin{document}
\newcommand{\beq}{\begin{equation}}
\newcommand{\beqa}{\begin{eqnarray}}
\newcommand{\eeq}{\end{equation}}
\newcommand{\eeqa}{\end{eqnarray}}
\newcommand{\non}{\nonumber}
\newcommand{\lb}{\label}
\newcommand{\fr}[1]{(\ref{#1})}
\newcommand{\cc}{\mbox{c.c.}}
\newcommand{\nr}{\mbox{n.r.}}
\newcommand{\bC}{\mbox{\boldmath {$C$}}}
\newcommand{\bp}{\mbox{\boldmath {$p$}}}
\newcommand{\bx}{\mbox{\boldmath {$x$}}}
\newcommand{\bF}{\mbox{\boldmath {$F$}}}
\newcommand{\bDelta}{\mathbf{\Delta}}
\newcommand{\ve}{{\varepsilon}}
\newcommand{\e}{\mbox{e}}
\newcommand{\tA}{\widetilde A}
\newcommand{\tB}{\widetilde B}
\newcommand{\tC}{\widetilde C}
\newcommand{\tU}{\widetilde U}
\newcommand{\tv}{\widetilde v}
\newcommand{\ts}{\widetilde s}
\newcommand{\ud}{\underline{\delta}}
\newcommand{\uD}{\underline{\Delta}}
\newcommand{\chN}{\check{N}}
\newcommand{\ucPhi}[1]{\TYPE 1 {\underline{\check{\Phi}}_{#1}} {}}
\newcommand{\ucPsi}[1]{\TYPE 1 {\underline{\check{\Psi}}_{#1}} {}}
\newcommand{\uctPhi}[1]{\TYPE 1 {\underline{\check{\Phi}}^{T}_{#1}} {}}
\newcommand{\uctPsi}[1]{\TYPE 1 {\underline{\check{\Psi}}^{T}_{#1}} {}}
\newcommand{\cA}{{\cal A}}
\newcommand{\cM}{{\cal M}}
\newcommand{\cU}{{\cal U}}
\newcommand{\cV}{{\cal V}}
\newcommand{\tcA}{\widetilde{\cal A}}
\newcommand{\DD}{{\cal D}}
\newcommand\TYPE[3]{ \underset {(#1)}{\overset{{#3}}{#2}}  }
\newcommand{\bfe}{ \TYPE 1    {\mathbf e}  {}   }
\newcommand{\bfb}{ \TYPE 1 {\mathbf b}   {}}
\newcommand{\bfh}{ \TYPE 1 {\mathbf h}   {}}
\newcommand{\bfj}{ \TYPE 1 {\mathbf j}   {}}
\newcommand{\bfn}{ \TYPE 1 {\mathbf n}   {}}
\newcommand{\bfA}{ \TYPE 1 {\mathbf A}   {}}
\newcommand{\bfB}{ \TYPE 2 {\mathbf B}   {}}
\newcommand{\bb}{\mbox{\boldmath {$b$}}}
\newcommand{\br}{\mbox{\boldmath {$r$}}}
\newcommand{\bn}{\mbox{\boldmath {$n$}}}
\newcommand{\bt}{\mbox{\boldmath {$t$}}}
\newcommand{\A}{   \TYPE 1  {A}  {}   }
\newcommand{\Ap}{  \TYPE p  {A}  {}   }
\newcommand{\cAone}{   \TYPE 1  { {\cal A} }  {}   }
\newcommand{\ucA}{ \TYPE 1 {\underline{\cal A}} {}   }
\newcommand{\F}{   \TYPE 2  {F}  {}   }
\newcommand{\cF}{   \TYPE 2  {\cal F}  {}   }
\newcommand{\Jtwo}{  \TYPE 2  {J}  {}   }
\newcommand{\cS}{  \TYPE m  {\check{S}} {}   }
\newcommand{\Lam}{ \TYPE q  {\Lambda}   {}   }
\newcommand{\alp}{ \TYPE p  {\alpha}   {}   }
\newcommand{\bep}{ \TYPE p  {\beta}   {}   }
\newcommand{\hash}{\#}
\newcommand{\hashat}{\widehat{\#}}
\newcommand{\D}{{\boldsymbol d}\,}
\newcommand\NN[1]{{\cal N}_{#1}}
\newcommand\MM[1]{{\cal M}_{#1}}
\newcommand\BAE[1]{{\begin{equation}{\begin{aligned}#1\end{aligned}}\end{equation}}}
\newcommand{\GamCLamM}[1]{{\Gamma\mathbb{C}\Lambda^{{#1}\cal{U}}}}
\newcommand{\normM}[2]{\left(  #1\, , \, #2 \right)}
\newcommand{\normU}[2]{\left\{ #1\, , \, #2 \right\}}
\def\man{{M}}
\def\GamLamM#1{{\Gamma \Lambda^{#1}\,\cal{U}}}
\newcommand{\res}{\mbox{res}}
\title{Amplitude equations \\
for a linear wave equation in a weakly curved pipe}
\author{ Shin-itiro GOTO\\
Physics Department, Lancaster University, Lancaster, LA1 4YB, UK.
}
\date{\today}
\maketitle
\begin{abstract}%
We study boundary effects in 
a linear wave equation with Dirichlet type 
conditions in a weakly curved pipe. 
The coordinates in our pipe are prescribed by a given small curvature with
finite range, while 
the pipe's cross section being circular. 
Based on the straight pipe case a perturbative analysis by which 
the boundary value conditions are exactly satisfied is employed.
As such an analysis we decompose the wave equation into 
a set of ordinary differential equations perturbatively.
We show the conditions when secular terms due to the curbed boundary 
appear in the naive
peturbative analysis. 
In eliminating such a secularity with a
singular perturbation method, we derive amplitude equations
and show that the eigenfrequencies in time are shifted due to the curved boundary.

\end{abstract}%
%
\section{Introduction}
In a spatially extended system a perfunctory theoretical description to  
explain experimentally observed  phenomena  
could be done by ignoring or simplifying boundary conditions. 
However in reality, the
observed phenomena in finite domain are generally affected by spatial boundary
conditions and are non-trivial. 
A reason why boundary conditions are ignored or simplified is 
the lack of mathematical tools. It has
been difficult to study such boundary effects in a systematic manner. 
We then need to develop a theoretical framework
that enables us to describe some effects of spatial boundary
conditions.  A first attempt might rely on a perturbative analysis,
and the small parameter is related to the magnitude of deformation
from trivial spatial boundary. 
In dynamical systems theory there are a variety of useful methodologies for 
dealing with perturbed systems\cite{GH}. 
With those methodologies differential
geometry provides powerful mathematical tools 
for it\cite{Nak90}.    
Beyond some existing works on this context, such as electromagnetic waves
in a curved pipe\cite{T07,GT09}, 
quantum eigenstates of a curved nanowire\cite{GW06}, one would like to 
know the higher order corrections in the magnitude of deformation of
boundaries.
To clearly see what can be observed including higher orders due to non-trivial 
spatial boundary conditions one needs to have a simple model.
As such a model, we consider a linear wave equation which is widely studied
in physical sciences, for example, fluid dynamics, electromagnetism 
and high energy physics.
In this paper the linear wave equation, or strictly speaking 
the classical complex Klein-Gordon equation,  with
Dirichlet type conditions is studied based on the language of
differential geometry with the use of perturbation methods.
There we will derive amplitude equations. 
In fact
the idea of amplitude equations is often used 
in order to study weakly perturbed systems in nonlinear science and
give benefits.  
We will see that even in the linear equation a mode-coupling
phenomenon occurs due to the weakly deformed boundary, and then 
perturbative correction terms for the mode-amplitude 
will be obtained using a singular perturbation method.

\section{Coordinate system, co-frame and Laplacians }
\label{sec:coodinate}
\begin{figure}[b]
\unitlength 1mm
\begin{picture}(120,30)
\put(50,0){\includegraphics[width=6.5cm]{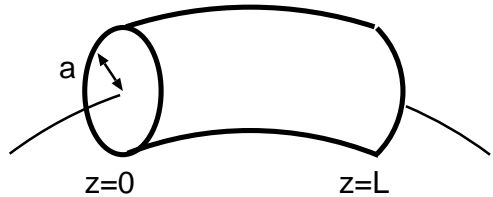}}
\end{picture}
\caption{Curved pipe}
\label{fig:curved-pipe}
\end{figure}

First we give the expression for the coordinate system adapted to our
curved pipe. 
The $z$-coordinate for our curved pipe which we consider is assumed to
be planer and prescribed by the curvature which is small.
Accordingly we do not consider the spatial curve, and our curve is 
nearly straight.  This assumption on curvature makes 
the perturbative analysis effective. 
The two-dimensional cross section at any $z$ in our pipe 
is assumed to be circle whose radius is $a$ ( See Fig.\ref{fig:curved-pipe} ). 

We denote the given curvature by 
\beq
\kappa(z)=\epsilon\kappa_0(z),
\label{eqn:kappa}
\eeq
where $|\kappa(z)a|\ll 1$ and $\epsilon$ is the small parameter.

To describe the geometry of boundary we use the following metric tensor 
$$
g= e^1\,\otimes\, e^1 
+ e^2\,\otimes\, e^2 + e^3\,\otimes\, e^3,
$$ 
where $\{e^1,\,e^2,\,e^3\}$ is the co-frame to be specified later, 
and our space will be denoted by $\cU$.
The canonical volume elements are defined as 
$$
\hash 1 = e^1\wedge e^2 \wedge e^3,\qquad 
\hashat 1= e^1\wedge e^2.
$$
Then the Hodge maps $\hash,\,\hashat$ are defined :
$$
\hash : \GamLamM{q}\to\GamLamM{3-q},\qquad 
\hashat : \GamLamM{q}\to\GamLamM{2-q},
$$
acting on 
$\{\,e^1,e^2,\e^3\,\}$ and 
$\{\,e^1,\e^2\,\}$, respectively.
Here $\GamLamM{q}$ denotes the set of $q$-form fields on $\cU$.
%

For our weakly curved pipe the adapted co-frame is derived from the 
use of Frenet frame. The space-curve is given in terms of Euclidean
position vector $\bC(z)$ then the points in the interior can be 
written as 
$$
\br(z,r,\theta)=\bC (z)+x_1(r,\theta)\,\bn(z) +x_2(r,\theta)\,\bb(z).
$$
Here $x_1$ and $x_2$ can be chosen as 
$x_1=r\cos\theta, x_2=r\sin\theta$ with $0\leq r\leq a$ and $0\leq\theta<2\pi$.
In addition, 
$\bn$ is the Frenet normal and $\bb$ is the Frenet bi-normal
vectors to the curve, the tangent vector $\bt$ is given by $d\bC /dz$.
The relations between them are describes as the Frenet-Serret formulae:  
$$
\frac{d}{dz}\bC=\bt,
\qquad \frac{d}{dz}\bt=\kappa\bn,
\qquad \frac{d}{dz}\bn=-\kappa\bt+\tau\bb,
\qquad \frac{d}{dz}\bb=-\tau\bn,
$$
with $\tau(z)$ being the Frenet torsion, which is zero in our case. 
Since the infinitesimal deviation of $\br$ is written with 
$\delta x_1,\delta x_2$ and $\delta z$ that are infinitesimal
deviations of $x_1,x_2$ and $z$, 
$$
\delta\br=\bn(\delta x_1-x_2\tau \delta z)+
\bb(\delta x_2+x_1\tau\delta z)+ \bt(1-\kappa x_1)\delta z,
$$
one can choose a convenient  orthonormal co-frame with $\tau=0$
for the interior domain
$$
\{\qquad  e^1=dr\qquad e^2=rd\theta,\qquad
e^3=(1-\epsilon\kappa_0(z)r\cos\theta) dz\qquad \},
$$
Taking 
the limit $\epsilon\to 0$ above
one has the adapted co-frame for the straight pipe 
$$
\{\qquad  e^1=dr\qquad e^2=rd\theta,\qquad e^3=dz\qquad \}.
$$
These are the usual cylindrical ones.
\subsection{Laplacians}
The Laplacian on $\cU$ 
for differentiable functions is 
\beq
\hash\, d\,\hash\, d,
\label{eqn:g-laplacian}
\eeq
where $d:\GamLamM{p}\to\GamLamM{p+1}$ is the exterior differentiation
in a patch with coordinates in $\cU$. So, for any function of $t,z,r,\theta$
one has 
$$
df(t,z,r,\theta)=\frac{\partial f}{\partial r}dr
+\frac{\partial f}{r\partial\theta}d\theta
+\frac{\partial f}{\partial z}dz.
$$
The explicit form of \fr{eqn:g-laplacian} for functions is calculated to be 
\BAE{
\hash\,d\,\hash\,d=
\left\{\frac{1}{(1-\kappa(z)r\cos\theta)^2}
\frac{\partial^2}{\partial z^2}
+\frac{\kappa'(z)r\cos\theta}{(1-\kappa(z) r\cos\theta)^3}\frac{\partial}{\partial z}
\right\}\non\\
+\left\{\frac{\partial^2}{\partial r^2}+\frac{\partial}{r\partial r}
-\frac{\kappa(z)\cos\theta}{1-\kappa(z)r\cos\theta}\frac{\partial}{\partial
r}\right\}
+\left\{\frac{\partial^2}{r^2\partial\theta^2}
+\frac{\kappa(z)\sin\theta}{r(1-\kappa(z)r\cos\theta)}
\frac{\partial}{\partial\theta}
\right\},
\non
}
where ${}'$ denotes the differentiation with respect to $z$.
When $\kappa=0$ this expression of the Laplacian reduces to 
\BAE{
\triangle^{(0)}:=\hash\,d\,\hash\,d\bigg|_{\epsilon=0}=
\frac{\partial^2}{\partial z^2}+\frac{\partial^2}{\partial r^2}
+\frac{\partial}{r\partial r}+\frac{\partial^2}{r^2\partial\theta^2}.
\label{eqn:Tri0}
}
Higher order corrections are obtained by substituting 
$\kappa(z)=\epsilon\kappa_0(z)$, \fr{eqn:kappa}, and expanding it in
$\epsilon$ as  
$$
\hash d\hash d = \triangle^{(0)}+\epsilon\triangle^{(1)}
+\epsilon^2\triangle^{(2)}+\cdots, 
$$ 
where 
\beqa
\triangle^{(1)}&:=&2\kappa_0(z) r\cos\theta\frac{\partial^2}{\partial z^2}
+\kappa_0'(z)r\cos\theta\frac{\partial}{\partial z}-
\kappa_0(z)\cos\theta\frac{\partial}{\partial r}
+\kappa_0(z)\frac{\sin\theta}{r}\frac{\partial}{\partial \theta},
\non\\
\triangle^{(2)}&:=&
3(\kappa_0(z)r\cos\theta)^2\frac{\partial^2}{\partial z^2}
+3\kappa_0(z)\kappa_0'(z)(r\cos\theta)^2\frac{\partial}{\partial z}
-(\kappa_0(z)\cos\theta)^2r\frac{\partial}{\partial r}
+\kappa_0^2(z)\sin\theta\cos\theta\frac{\partial}{\partial\theta}.
\non
\eeqa

Similarly the two-dimensional Laplacian for any function of
$(r,\theta)$ 
is generalized to
$$
\hashat\,d\,\hashat\,d,
$$
The perturbation due to the curvature does not affect the 
two-dimensional Laplacian, and then the explicit form of this 
is calculated as 
$$
\hashat\,d\,\hashat\,d
=\frac{\partial^2}{\partial r^2}+\frac{\partial}{r\partial r}
+\frac{\partial^2}{r^2\partial\theta^2 }.
$$
\subsection{Eigenvalue problems of Laplacians at $\epsilon=0$}
The eigenvalue problem associated with the two-dimensional 
Laplacian, acting on functions of 
$(r,\theta)$ that satisfy the Dirichlet condition $\Phi_N(a,\theta)=0$, 
is written as  
$$
\hashat\,d\,\hashat\,d\, \Phi_N(r,\theta)=-\beta_N^2\Phi_N(r,\theta).
$$
An explicit form of the solution $\Phi_N$ is found to be 
$$
\Phi_N(r,\theta)=J_n\left(\frac{x_{q(n)}}{a}r\right)\,e^{in\theta},
$$
and the constant $\beta_N$ is 
$$
\beta_N=\frac{x_{q(n)}}{a},\qquad J_n(x_{q(n)})=0, 
$$
with $N=\{n,q(n)\}, n=0,\pm 1,\pm 2,\dots,$ and $q=1,2,\dots$, 
$J_n$ is the $n$-th order Bessel function and $x_{q(n)}$ is the $q$-th 
zero of $J_n$.
The orthogonality can be shown as 
\beq
\int_\DD\overline{\Phi_M}\Phi_N\hashat 1=\NN{N}^2\delta_{N,M}.
\label{eqn:orthogonality}
\eeq
where 
$$
\DD:=\{\quad (r,\theta)\quad |\quad 0\leq r \leq a, \quad 
0\leq\theta< 2\pi\quad \},
$$
$$
\NN{N}^2=\pi a^2 J_{n+1}^2 (x_{q(n)}),\qquad 
\delta_{N,M}:=\delta_{n,m}\delta_{q,p}
$$ 
with $N=\{n,q(n)\}, \,M=\{m,p(m)\}$ and $\delta_{a,b}$ being the
Kronecker delta symbol. 

The eigenvalue problem associated with the three-dimensional 
Laplacian at $\epsilon=0$, acting on functions of 
$(z,r,\theta)$ that satisfy Dirichlet condition 
$\varphi_{N,\eta}=0$ at boundary, 
is solved as 
$$
\triangle^{(0)}\varphi_{N,\eta}(z,r,\theta)
=-\left\{\beta_N^2+\left(\frac{\eta\pi}{L}\right)^2\right\}
\varphi_{N,\eta}(z,r,\theta),\qquad 
\varphi_{N,\eta}(z,r,\theta)=\Phi_N(r,\theta)\sin\left(\frac{\eta\pi}{L}z\right).
$$
Here $\eta=1,2,\dots$ and the boundary is located at 
$$
z=0,L,\qquad r=a.
$$
The eigenfunction $\varphi_{N,\eta}$ also satisfies 
\beq
\left(\triangle^{(0)}+\mu^2\right)\varphi_{N,\eta}(z,r,\theta)
=-\left\{\beta_N^2+\left(\frac{\eta\pi}{L}\right)^2+\mu^2\right\}
\varphi_{N,\eta}(z,r,\theta),
\label{eqn:eigenvalue-problem}
\eeq
with the same boundary conditions where $\mu$ is a  constant.
\section{The wave equation in the curved pipe 
and its naive perturbative analysis}
We study the following complex-valued linear partial differential equation
for $\phi(\epsilon,t,z,r,\theta)\in\mathbb{C}$ 
\beq
\left\{\frac{1}{c^2}\frac{\partial^2}{\partial t^2}-
\hash\,d\,\hash\,d +\mu^2
\right\}\phi(\epsilon,t,z,r,\theta)=0,
\label{eqn:basic-pde}
\eeq
where $c,\mu$ are constants, the operators appeared here are 
defined in \S\ref{sec:coodinate}, and 
the ranges of coordinates are 
$$
-\infty \leq t\leq\infty,\qquad 
0\leq z\leq L,\qquad 
0 \leq  r\leq\ a,\qquad 
0 \leq  \theta< 2\pi. 
$$
The boundary condition is 
\beq
\phi(\epsilon,t,z,r,\theta)|_{\partial{\cal U}} =0,
\label{eqn:bd-condition}
\eeq
where $\partial{\cal U}$ denotes spatial boundary.
In this paper \fr{eqn:basic-pde} is referred to as the wave equation.

To obtain an approximate solution 
we expand 
the solution based on the lowest order problem
\BAE{
\phi(\epsilon,t,z,r,\theta)
=\sum_{\eta=1}^{\infty}\sum_N
\Phi_N(r,\theta)\sin\left(\frac{\eta\pi}{L}z\right)
A_{N,\eta}(\epsilon,t),\\ 
A_{N,\eta}(\epsilon,t)
= A_{N,\eta}^{(0)}(t)+\epsilon A_{N,\eta}^{(1)}(t)
+\epsilon^2 A_{N,\eta}^{(2)}(t)+{\cal O}(\epsilon^3),\\
\label{eqn:phi-expand-L}
}
and correspondingly
\beqa
\phi(\epsilon,t,z,r,\theta)&=&
\phi^{(0)}(t,z,r,\theta)
+\epsilon\phi^{(1)}(t,z,r,\theta)
+\epsilon^2\phi^{(2)}(t,z,r,\theta)+{\cal O}(\epsilon^3),\non\\
\phi^{(j)}(t,z,r,\theta)&:=&
\sum_{\chN}\Phi_N(r,\theta)\sin\left(\frac{\eta\pi}{L}z\right)
A_{\chN}^{(j)}(t),\qquad j=0,1,2,\dots,
\non
\eeqa
where $\chN:=\{N,\eta\}$ which we call a mode. 
This expansion of the solution in $\epsilon$  
and $A_{\chN}$ are refereed to as the naive
perturbation expansion and the mode-amplitude associated with the mode
$\chN$ respectively. Note here that this form of the expansion of the solution 
satisfies the boundary conditions, \fr{eqn:bd-condition}, 
for any value of $\epsilon$.

To obtain equations for $A_{\chN}^{(j)}$ from those for
$\phi^{(j)}$, one uses the following identity 
\beq
A_{\chN}^{(j)}(t)=
\frac{1}{\NN{N}^2}\int_{\DD}\overline{\Phi_N}\hashat 1\frac{2}{L}\int_0^Ldz\, 
\sin\left(\frac{\eta\pi}{L}z \right)\,
\phi^{(j)}(t,z,r,\theta).
\label{eqn:A-from-phi}
\eeq

\subsection{Unperturbed solution}
To zeroth order in $\epsilon$ the equation of motion becomes
\beq
\left\{\frac{1}{c^2}\frac{\partial^2}{\partial t^2}-\triangle^{(0)}
+\mu^2\right\}\phi^{(0)}(t,z,r,\theta)=0,
\label{eqn:unperturbed-phi}
\eeq
where $\triangle^{(0)}$ is defined in \fr{eqn:Tri0}. This can be written in
terms of $A_{\chN}^{(0)}$ using \fr{eqn:eigenvalue-problem} and 
\fr{eqn:A-from-phi} as 
$$
\ddot{A}_{\chN}^{(0)}+\Omega_{\chN}^2 A_{\chN}^{(0)}=0,
\qquad
\Omega_{\chN}:=c\,\sqrt{\left(\frac{\eta\pi}{L}\right)^2+\beta_N^2
+\mu^2},
$$
where $\dot{}$ denotes the differentiation with respect to $t$.
The solution is  
\beq
A_{\chN}^{(0)}(t)=A_{\chN}^{(0,-)}\,e^{-i\Omega_{\chN}t}
+A_{\chN}^{(0,+)}\,e^{i\Omega_{\chN}t},\qquad 
\label{eqn:A0-sol}
\eeq
where $A_{\chN}^{(0,\pm)}\in\mathbb{C}$ are integral constants.

\subsection{First order solution}
To first order in $\epsilon$ the equation of motion is written as 
\beqa
&&\left\{\frac{1}{c^2}\frac{\partial}{\partial t^2}-\triangle^{(0)}
+\mu^2\right\}\phi^{(1)}\non\\
&&\qquad =
2\kappa_0(z) \phi^{(0)\prime\prime} r\cos\theta
+\kappa_0'(z) \phi^{(0)\prime} r\cos\theta
-\kappa_0(z)\frac{\partial\phi^{(0)}}{\partial r}\cos\theta
+\kappa_0(z)\frac{\partial\phi^{(0)}}{\partial\theta}\frac{\sin\theta}{r},
\non
\eeqa
or equivalently,
\beq
\ddot{A}_{\chN}^{(1)}+\Omega_{\chN}^2 A_{\chN}^{(1)} 
=\frac{2}{L}\left(\frac{c}{\NN{N}}\right)^2\sum_{\chN'}F_{\chN,\chN'}^{(1)}
A_{\chN'}^{(0)}(t).
\label{eqn:A1}
\eeq
Here
$$
F_{\chN,\chN'}^{(1)}:= \left\{-2\left(\frac{\eta'\pi}{L}\right)^2
\langle\kappa_0\rangle_{S_\eta,S_{\eta'}}
+\frac{\eta'\pi}{L}\langle\kappa_0'\rangle_{S_\eta,C_{\eta'}}\right\}C_{N,N'}^{(1,1)}
+\langle\kappa_0\rangle_{S_\eta,S_{\eta'}}C_{N,N'}^{(1,2)},\quad\in\mathbb{R}
$$
\beqa
C_{N,M}^{(1,1)}&:=&\int_\DD\overline{\Phi_N}\Phi_M r\cos\theta\hashat 1
=\pi(\delta_{m,n-1}+\delta_{m,n+1})\int_0^adr r^2
J_n\left(\frac{x_{q(n)}}{a}r\right)
J_m\left(\frac{x_{p(m)}}{a}r\right),\quad\in\mathbb{R}
\non\\
C_{N,M}^{(1,2)}&:=&\int_\DD\overline{\Phi_N}
\left(
\frac{\partial\Phi_M}{\partial\theta}\frac{\sin\theta}{r}-
\frac{\partial\Phi_M}{\partial r}\cos\theta
\right)\hashat 1\non\\
&=&\pi m(\delta_{m,n-1}-\delta_{m,n+1}) 
\int_0^a drJ_n\left(\frac{x_{q(n)}}{a}r\right)
J_m\left(\frac{x_{p(m)}}{a}r\right) \non\\
&&\qquad -\pi\frac{x_{p(m)}}{a}(\delta_{m,n-1}+\delta_{m,n+1})
\int_0^a dr r J_n\left(\frac{x_{q(n)}}{a}r\right) 
J_m'\left(\frac{x_{p(m)}}{a}r\right).\quad\in\mathbb{R}\non
\eeqa
In addition, for a function $f$ of $z$, we have defined
$$
\langle f\rangle_{S_\eta,S_{\eta'}}:=
\int_0^L dz\,\sin\left(\frac{\eta\pi}{L}z\right)
\sin\left(\frac{\eta'\pi}{L}z\right)f(z),\quad
\langle f\rangle_{S_\eta,C_{\eta'}}:=
\int_0^L dz\,\sin\left(\frac{\eta\pi}{L}z\right)
\cos\left(\frac{\eta'\pi}{L}z\right)f(z).
$$
Due to $\delta_{n',n\pm1}$ in $C_{N,N'}^{(1,1)}$ and $C_{N,N'}^{(1,2)}$
the right hand side of \fr{eqn:A1} reduces to 
$$
\frac{2}{L}\left(\frac{c}{\NN{N}}\right)^2\sum_{\chN'}F_{\chN,\chN'}^{(1)}
A_{\chN'}^{(0)}(t)=\frac{2}{L}\left(\frac{c}{\NN{N}}\right)^2
\sum_{\eta'}\sum_{n'=n\pm 1}\sum_{q'(n')}
F_{\chN,n'q'(n'),\eta'}^{(1)}
A_{n',q'(n'),\eta'}^{(0)}(t).
$$

After substituting
\fr{eqn:A0-sol}  into \fr{eqn:A1} 
one can find the solution (See \S\ref{appendix})
\beqa
A_{\chN}^{(1)}(t)&=&
\frac{2}{L}\left(\frac{c}{\NN{N}}\right)^2
\sum_{\eta'} \sum_{n'=n\pm 1} \sum_{q'(n')}\Bigg\{
R_{\chN,\chN'}F_{\chN,n',q'(n'),\eta'}^{(1)}t\left(
 \frac{A_{\chN'}^{(0,-)}}{-2i\Omega_{\chN}}\,e^{-i\Omega_{\chN}t}
+\frac{A_{\chN'}^{(0,+)}}{2i\Omega_{\chN}}\,e^{i\Omega_{\chN}t}
\right)
\non\\
&&+
N_{\chN,\chN'}F_{\chN,n',q'(n'),\eta'}^{(1)}
\frac{A_{\chN'}^{(0,-)}\,e^{-i\Omega_{\chN'}t}+A_{\chN'}^{(0,+)}\,e^{i\Omega_{\chN'}t}
}{\Omega_{\chN}^2-\Omega_{\chN'}^2 }
\Bigg\},
\label{eqn:A1-sol}
\eeqa
where $R_{\chN,\chN'}$ and $N_{\chN,\chN'}$ are defined as 
$$
R_{\chN,\chN'}:=\delta_{\Omega_{\chN},\Omega_{\chN'}},\qquad 
N_{\chN,\chN'}:=1-\delta_{\Omega_{\chN},\Omega_{\chN'}}.
$$
The condition $R_{\chN,\chN'}=1$ at this order 
is equivalent to 
the resonance conditions
\beq
\Omega_{n,q(n),\eta}=\Omega_{n\pm 1,q'(n\pm 1),\eta'}.
\label{eqn:res-pm1}
\eeq
On the other hand the condition $N_{\chN,\chN'}=1$  at this
order is the case when 
$$
\Omega_{n,q(n),\eta}\neq\Omega_{n\pm 1,q'(n\pm 1),\eta'}.
$$ 
Note that there are secular terms, $\propto t$, in \fr{eqn:A1-sol} 
when the resonance condition is satisfied, and that 
the resonance condition, \fr{eqn:res-pm1}, 
does not contain $\kappa(z)$.

\subsection{Second order solution}
\label{sec:second-order}
To second order in $\epsilon$ the equation of motion is written as 
\beqa
&&\left\{\frac{1}{c^2}\frac{\partial^2}{\partial t^2}-\triangle^{(0)}
+\mu^2\right\}\phi^{(2)}\non\\
&&=
2\kappa_0(z) \phi^{(1)\prime\prime} r\cos\theta
+\kappa_0'(z) \phi^{(1)\prime} r\cos\theta
-\kappa_0(z)\frac{\partial\phi^{(1)}}{\partial r}\cos\theta
+\kappa_0(z)\frac{\partial\phi^{(1)}}{\partial\theta}\frac{\sin\theta}{r}\non\\
&& 
+3(\kappa_0(z)r\cos\theta)^2\phi^{(0)\prime\prime}+3\kappa_0(z)\kappa_0'(z)
(r\cos\theta)^2\phi^{(0)\prime}
-\kappa_0^2(z)r\frac{\partial\phi^{(0)}}{\partial r}\cos^2\theta
+\kappa_0^2(z)\frac{\partial\phi^{(0)}}{\partial\theta}\cos\theta\sin\theta
,
\non
\eeqa
or equivalently,
\beq
\ddot{A}_{\chN}^{(2)}+\Omega_{\chN}^2 A_{\chN}^{(2)} 
=\frac{2}{L}\left(\frac{c}{\NN{N}}\right)^2
\sum_{\eta'}\left\{\sum_{n'=n\pm 1}\sum_{q'(n')} F_{\chN,\chN'}^{(1)}
A_{\chN'}^{(1)}(t)+
\sum_{n'=n,n\pm 2}\sum_{q'(n')} F_{\chN,\chN'}^{(2)}
A_{\chN'}^{(0)}(t)\right\}.
\label{eqn:A2}
\eeq
Here
$$
F_{\chN,\chN'}^{(2)}:=\left\{
-3\left(\frac{\eta'\pi}{L}\right)^2\langle\kappa_0^2\rangle_{S_\eta,S_{\eta'}}
+3\frac{\eta'\pi}{L}\langle\kappa_0\kappa_0'\rangle_{S_\eta,C_{\eta'}}\right\}
C_{N,N'}^{(2,1)}
-\langle\kappa_0^2\rangle_{S_\eta,S_{\eta'}}C_{N,N'}^{(2,2)}
+\frac{i}{2}n'\langle\kappa_0^2\rangle_{S_\eta,S_{\eta'}}C_{N,N'}^{(2,3)},
$$
with 
\beqa
C_{N,N'}^{(2,1)}&:=&
\int_\DD\overline{\Phi_N}\Phi_{N'}r^2\cos\theta\hashat 1\non\\
&=&\pi\left\{\delta_{n',n}+\frac{1}{2}(\delta_{n',n+2}+\delta_{n',n-2})
\right\}
\int_0^a
dr\,r^3J_n\left(\frac{x_{q(n)}}{a}r\right)
J_{n'}\left(\frac{x_{q'(n')}}{a}r\right),\quad\in\mathbb{R}
\non\\
C_{N,N'}^{(2,2)}&:=&
\int_\DD\overline{\Phi_N}\frac{\partial\Phi_{N'}}{\partial r}
r\cos^2\theta\hashat 1\non\\
&=&\pi\left\{\delta_{n',n}+\frac{1}{2}(\delta_{n',n+2}+\delta_{n',n-2})
\right\}\frac{x_{q'(n')}}{a}\int_0^adr\,r^2 
J_n\left(\frac{x_{q(n)}}{a}r\right)
J'_{n'}\left(\frac{x_{q'(n')}}{a}r\right),\quad\in\mathbb{R}\non\\
C_{N,N'}^{(2,3)}&:=&
\int_\DD\overline{\Phi_N}\Phi_{N'}\sin2\theta\hashat 1\non\\
&=&-\pi i (\delta_{n',n-2}-\delta_{n',n+2})
\int_0^a dr r J_n\left(\frac{x_{q(n)}}{a}\right) 
J_{n'} \left(\frac{x_{q'(n')}}{a}\right).
\quad\in\,i\,\mathbb{R}\non
\eeqa

One obtains the equation of motion at this order after substituting
the lower solutions $A^{(0)}_{\chN}(t)$ and 
$A^{(1)}_{\chN} (t)$, given in \fr{eqn:A0-sol} and \fr{eqn:A1-sol}, 
 into \fr{eqn:A2}.
The right hand side of \fr{eqn:A2} can then be expressed as 
\beqa
&&\frac{2}{L}\left(\frac{c}{\NN{N}}\right)^2
\sum_{\eta'}\sum_{n'=n\pm 1}\sum_{q'(n')} F_{\chN,\chN'}^{(1)}
\bigg[
\frac{2}{L}\left(\frac{c}{\NN{N'}}\right)^2
\sum_{\eta''} \sum_{n''=n'\pm 1} \sum_{q''(n'')}\non\\
&&\quad\times\bigg\{
R_{\chN',\chN''}F_{\chN',\chN''}^{(1)}
t\left(\frac{A_{\chN''}^{(0,-)}}{-2i\Omega_{\chN'}}
e^{-i\Omega_{\chN'}t}
+
\frac{A_{\chN''}^{(0,+)}}{2i\Omega_{\chN'}}
e^{i\Omega_{\chN'}t}\right)\non\\
&&\quad
+
N_{\chN',\chN''}
\frac{F_{\chN',\chN''}^{(1)}}{\Omega_{\chN'}^2-\Omega_{\chN''}^2 }
\left(A_{\chN''}^{(0,-)}\,e^{-i\Omega_{\chN''}t}
+A_{\chN''}^{(0,+)}\,e^{i\Omega_{\chN''}t}\right)
\bigg\}%
\bigg]
\non\\
&&+\frac{2}{L}\left(\frac{c}{\NN{N}}\right)^2
\sum_{\eta''}\sum_{n''=n,n\pm 2}\sum_{q''(n'')} F_{\chN,\chN''}^{(2)}
\left(
A_{\chN''}^{(0,-)}\,e^{-i\Omega_{\chN,\chN''}t}
+A_{\chN''}^{(0,+)}\,e^{i\Omega_{\chN,\chN''}t}
\right),
\non
\eeqa
from which one has the solution 
\beqa
&&A_{\chN}^{(2)}(t)=
\frac{2}{L}\left(\frac{c}{\NN{N}}\right)^2
\sum_{\eta'}\sum_{n'=n\pm 1}\sum_{q'(n')} F_{\chN,\chN'}^{(1)}
\frac{2}{L}\left(\frac{c}{\NN{N'}}\right)^2
\sum_{\eta''} \sum_{n''=n'\pm 1} \sum_{q''(n'')}\non\\
&&\quad\times\bigg[
R_{\chN',\chN''}
\frac{F_{\chN',\chN''}^{(1)}}{-2i\Omega_{\chN'} }
A_{\chN''}^{(0,-)}
\bigg\{
R_{\chN,\chN''}\bigg(
\frac{it^2}{4\Omega_{\chN}}+\frac{t}{4\Omega_{\chN}^2}
\bigg)
e^{-i\Omega_{\chN}t}+N_{\chN,{\chN''}}t
\frac{e^{-i\Omega_{\chN''}t}}{\Omega_{\chN}^2-\Omega_{\chN''}^2}\bigg\}\non\\
&&+
N_{\chN',{\chN''}}
\frac{F_{\chN',\chN''}^{(1)}}{\Omega_{\chN'}^2-\Omega_{\chN''}^2 }
A_{\chN''}^{(0,-)}\bigg\{R_{\chN,\chN''}\frac{t\,e^{-i\Omega_{\chN}t}}{-2i\Omega_{\chN}}+N_{\chN,\chN''}\frac{e^{-i\Omega_{\chN''}t}}{\Omega_{\chN}^2-\Omega_{\chN''}^2}\bigg\}
\non\\
&&
+R_{\chN',\chN''}
\frac{F_{\chN',\chN''}^{(1)}}{2i\Omega_{\chN'} }
A_{\chN''}^{(0,+)}
\bigg\{
R_{\chN,\chN''}\bigg(
\frac{-it^2}{4\Omega_{\chN}}+\frac{t}{4\Omega_{\chN}^2}
\bigg)
e^{i\Omega_{\chN}t}+N_{\chN,{\chN''}}t
\frac{e^{i\Omega_{\chN''}t}}{\Omega_{\chN}^2-\Omega_{\chN''}^2}\bigg\}\non\\
&&+
N_{\chN',{\chN''}}
\frac{F_{\chN',\chN''}^{(1)}}{\Omega_{\chN'}^2-\Omega_{\chN''}^2 }
A_{\chN''}^{(0,+)}\bigg\{R_{\chN,\chN''}\frac{t\,e^{i\Omega_{\chN}t}}{2i\Omega_{\chN}}+N_{\chN,\chN''}\frac{e^{i\Omega_{\chN''}t}}{\Omega_{\chN}^2-\Omega_{\chN''}^2}\bigg\}\bigg]\non\\
&&+\frac{2}{L}\left(\frac{c}{\NN{N}}\right)^2
\sum_{\eta''}\sum_{n''=n,n\pm 2}\sum_{q''(n'')} F_{\chN,\chN''}^{(2)}
\Bigg[
A_{\chN''}^{(0,-)}
\left\{
R_{\chN,\chN''}\frac{t\,e^{-i\Omega_{\chN}t}}{-2i\Omega_{\chN}}+N_{\chN,\chN''}
\frac{e^{-i\Omega_{\chN''}t}}{\Omega_{\chN}^2-\Omega_{\chN''}^2}\right\}\non\\
&&\qquad\qquad+
A_{\chN''}^{(0,+)}
\left\{
R_{\chN,\chN''}\frac{t\,e^{i\Omega_{\chN}t}}{2i\Omega_{\chN}}+N_{\chN,\chN''}
\frac{e^{i\Omega_{\chN''}t}}{\Omega_{\chN}^2-\Omega_{\chN''}^2}\right\}
\Bigg].
\label{eqn:A2-sol}
\eeqa
When the resonance condition at ${\cal O} (\epsilon)$ is  satisfied,  
the solution at ${\cal O}(\epsilon^2)$ includes the terms being
proportional to $t^2$. 
From \fr{eqn:A2-sol} the resonance conditions between 
different two modes are obtained as 
\beq
\Omega_{n,q(n),\eta}=\Omega_{n\pm 2,q''(n\pm 2),\eta''}. 
\label{eqn:res-pm2}
\eeq
Although when \fr{eqn:res-pm2} are not 
satisfied in addition to the case where \fr{eqn:res-pm1} are not satisfied, 
there are secular terms in $A_{\chN}^{(2)}(t)$ due to the self-mode couplings.
The existence of such secular behavior in
$A_{\chN}^{(2)}(t)$ is a notable qualitative 
difference from the analysis at ${\cal O}(\epsilon)$.
Such secular behavior is due to the summation ranges in  \fr{eqn:A2-sol} 
$$
\sum_{n''=n,n\pm 1},\qquad \mbox{and}\qquad
\sum_{n'=n\pm1}\sum_{n''=n'\pm 1}= \sum_{n'=n\pm1}\sum_{n''=n,n\pm 2}.
$$

Taking into account the resonance conditions up to 
${\cal O} (\epsilon^2)$ one rewrites \fr{eqn:A2-sol} explicitly 
in the following cases
\begin{itemize}
\item[(i)] 
\beqa
\Omega_{n,q(n),\eta}&\neq&\Omega_{n\pm 1,q'(n\pm 1),\eta'},\quad
\mbox{for all} \quad q'(n\pm 1),\eta'\qquad 
\mbox{and} \non\\
\Omega_{n,q(n),\eta}&\neq&\Omega_{n\pm 1,q''(n\pm 2),\eta''},\quad
\mbox{for all} \quad q''(n\pm 2),\eta'',\non
\eeqa
\item[(ii)]
\beqa
\Omega_{n,q(n),\eta}&\neq&\Omega_{n\pm 1,q'(n\pm 1),\eta'},\quad
\mbox{for all} \quad q'(n\pm 1),\eta',\qquad 
\mbox{and} \non\\
\Omega_{n,q(n),\eta}&=&\Omega_{n\pm 2,q_r''(n\pm 2),\eta_r''},\quad
\mbox{for some particular}\quad q_r''(n\pm 2),\eta_r'',
\label{eqn:case-ii-qr}
\eeqa
%
%
%
%
\item[(iii)]
\beqa
\Omega_{n,q(n),\eta}&=&\Omega_{n\pm 1,q_r'(n\pm 1),\eta_r'},\quad
\mbox{for some particular} \quad q_r'(n\pm 1),\eta_r',\qquad 
\mbox{and} 
\label{eqn:case-iii-qr-1}\\
\Omega_{n,q(n),\eta}&=&\Omega_{n\pm 2,q_r''(n\pm 2),\eta_r''},\quad
\mbox{for some particular} \quad q_r''(n\pm 2),\eta_r'',
\label{eqn:case-iii-qr-2}
\eeqa
\end{itemize}
Here the reason why the case 
\beqa
\Omega_{n,q(n),\eta}&=&\Omega_{n\pm 1,q_r'(n\pm 1),\eta_r'},\quad
\mbox{for some particular} \quad q_r'(n\pm 1),\eta_r',\qquad 
\mbox{and} 
\label{eqn:casse-no-appear-1}\\
\Omega_{n,q(n),\eta}&\neq&\Omega_{n,q''(n\pm 2),\eta''},\quad
\mbox{for all}\quad q''(n\pm 2),\eta'',
\label{eqn:case-no-appear-2}
\eeqa
does not appear is as follows.
If \fr{eqn:casse-no-appear-1} is satisfied, 
then one uses \fr{eqn:casse-no-appear-1} twice, 
and can find $q_r''(n\pm 2),\eta_r''$ that satisfy 
$$
\Omega_{n,q(n),\eta}=\Omega_{n,q_r''(n\pm 2),\eta_r''}.
$$
This is in contradiction to 
\fr{eqn:case-no-appear-2}.
\subsubsection{Case (i)}
\label{sec:second-order-case-i}
In the case, where both \fr{eqn:res-pm1} and \fr{eqn:res-pm2} are not
satisfied,  one has 
\beqa
&&A_{\chN}^{(2)}(t)=\frac{2}{L}\left(\frac{c}{\NN{N}}\right)^2
\Bigg[t
\left(\frac{A_{\chN}^{(0,-)}}{-2i\Omega_{\chN}}e^{-i\Omega_{\chN} t}
  +\frac{A_{\chN}^{(0,+)}}{2i\Omega_{\chN}}e^{i\Omega_{\chN} t}\right)
\non\\
&&\quad
\times\bigg\{
\sum_{\eta'}\sum_{n'=n\pm 1}\sum_{q'(n')}
F_{\chN,\chN'}^{(1)}\frac{2}{L}\left(\frac{c}{\NN{N'}}\right)^2
\frac{F_{\chN',\chN}^{(1)}}{\Omega_{\chN'}^2-\Omega_{\chN}^2}
+F_{\chN,\chN}^{(2)}\bigg\}\Bigg]+\,(\,\mbox{NS}\,),
\label{eqn:A2-sol-i}
\eeqa
where $F_{\chN,\chN}^{(2)}$ is real because 
$C_{n,q(n),n,q'(n)}^{(2,3)}\in i\,\mathbb{R}$ vanishes 
and NS is the abbreviation for the nonresonance terms.
The time-dependence of NS is $\exp(i\,\Omega_{\chN'}'t)$ with 
$\Omega_{\chN'}'\neq\Omega_{\chN}$ for a given $\chN$.

Thus when the resonance conditions between other different modes 
are not satisfied 
one observes a secular (divergent in time) behavior, 
$A_{\chN}^{(2)}(t)\propto t$.
Correspondingly,  our naive perturbative analysis could only be valid in
a short time range.  
To improve this naive
perturbative result one needs another perturbation method that will be
discussed in \S\ref{sec:amp-eqs}. 
\subsubsection{Case (ii)}
The solution in the case (ii) is written as 
\beqa
&&A_{\chN}^{(2)}(t)=\frac{2}{L}\left(\frac{c}{\NN{N}}\right)^2
\sum_{\eta'}\sum_{n'=n\pm 1}\sum_{q'(n')}
F_{\chN,\chN'}^{(1)}\frac{2}{L}
\left(\frac{c}{\NN{N'}}\right)^2
\non\\
&&\qquad
\times\,
\sum_{\eta''}^{(\res)}\sum_{n''=n'\pm 1}^{(\res)}\sum_{q''(n'')}^{(\res)}
\frac{F_{\chN',\chN''}^{(1)}}{\Omega_{\chN'}^2-\Omega_{\chN}^2}
t\left(\frac{ A_{\chN}^{(0,-)}}{-2i\Omega_{\chN}}
e^{-i\Omega_{\chN} t}
+\frac{ A_{\chN}^{(0,+)}}{2i\Omega_{\chN}}
e^{i\Omega_{\chN} t}\right)
\non\\
&&\quad
+\frac{2}{L}\left(\frac{c}{\NN{N}}\right)^2
\sum_{\eta''}^{(\res)}\sum_{n''=n'\pm 1}^{(\res)}\sum_{q''(n'')}^{(\res)}
F_{\chN,\chN''}^{(2)}
t\left(
\frac{A_{\chN''}^{(0,-)} }{-2i\Omega_{\chN}}e^{-i\Omega_{\chN}t}
+
\frac{A_{\chN''}^{(0,+)} }{2i\Omega_{\chN}}e^{i\Omega_{\chN}t}\right)
+\,(\,\mbox{NS}\,),
\label{eqn:A2-sol-ii}
\eeqa
where 
$$
\sum_{\eta''}^{(\res)}\sum_{n''=n'\pm 1}^{(\res)}\sum_{q''(n'')}^{(\res)}
$$
are the summations of $\eta'',n''$ and $q''$ that satisfy   
\fr{eqn:case-ii-qr} and $\Omega_{n,q(n),\eta}=\Omega_{n,q''(n),\eta''}$, 
the self-mode coupling, for a given set $\chN$. 


\subsubsection{Case (iii)}
The solution in the case (iii) is written as 
\beqa
&&A_{\chN}^{(2)}(t)=\frac{2}{L}\left(\frac{c}{\NN{N}}\right)^2
\sum_{\eta'}^{(\res)}\sum_{n'=n\pm 1}^{(\res)}\sum_{q'(n')}^{(\res)}
F_{\chN,\chN'}^{(1)}\frac{2}{L}
\left(\frac{c}{\NN{N'}}\right)^2
\non\\
&&
\times\,
\sum_{\eta''}^{(\res)}\sum_{n''=n'\pm 1}^{(\res)}\sum_{q''(n'')}^{(\res)}
F_{\chN',\chN''}^{(1)}\left\{
\frac{A_{\chN''}^{(0,-)}}{-2i\Omega_{\chN'}}
\left(\frac{it^2}{4\Omega_{\chN}}+\frac{t}{4\Omega_{\chN}^2}\right)
\, e^{-i\Omega_{\chN} t}
+
\frac{A_{\chN''}^{(0,+)}}{2i\Omega_{\chN'}}
\left(\frac{-it^2}{4\Omega_{\chN}}+\frac{t}{4\Omega_{\chN}^2}\right)
\, e^{i\Omega_{\chN} t}\right\}
\non\\
&&\quad
+\frac{2}{L}\left(\frac{c}{\NN{N}}\right)^2
\sum_{\eta''}^{(\res)}\sum_{n''=n'\pm 1}^{(\res)}\sum_{q''(n'')}^{(\res)}
F_{\chN,\chN''}^{(2)}t\left(
\frac{A_{\chN''}^{(0,-)}}{-2i\Omega_{\chN}}e^{-i\Omega_{\chN}t}
+\frac{A_{\chN''}^{(0,+)}}{2i\Omega_{\chN}}e^{i\Omega_{\chN}t}
\right)
+\,(\,\mbox{NS}\,),
\label{eqn:A2-sol-iv}
\eeqa 
where the summations are performed when \fr{eqn:case-iii-qr-1}, 
\fr{eqn:case-iii-qr-2} and 
$\Omega_{n,q(n),\eta}=\Omega_{n,q''(n),\eta''}$, 
the self-mode coupling, are satisfied for a given set $\chN$.
\section{Amplitude equations}
\label{sec:amp-eqs}
In this section, 
we derive amplitude equations by which long time behavior can
approximately be described for each case in \S\ref{sec:second-order}.

The procedure to obtain an amplitude equation will be shown  
in the case (i), where there is no resonance between different modes. 
Even in this case as mentioned in \S\ref{sec:second-order-case-i}, 
there is a secular term, $\propto\epsilon^2 t$ due to the self-mode
coupling. This does not make our
naive perturbative result valid globally in time, and
  valid only for  a short time range. 
To improve this one uses another perturbation method, rather than the
naive perturbation method. In this paper 
we use the renormalization method as a type of systematic perturbation method 
for removing secular terms caused by the use of
the naive perturbation method.
Using the renormalization method, one can obtain the equations of motion  
that describe the long-time behavior\cite{CGO},\cite{Kun},\cite{IN},\cite{Chi},\cite{GMN99}.  
In this paper we use the method in Ref.\cite{GMN99} 
by which reduced equations have been systematically obtained\cite{MN01}. 

For the other cases, (ii) and (iii) in \S\ref{sec:second-order}, 
the procedures to obtain amplitude
equations are similar to the case (i), and we then show only the resultant 
equations. 

\subsection{case (i)}
\label{sec:renorm-case-i}
In the case (i) one concentrates on   
\fr{eqn:A2-sol-i}, which contains secular terms due to the use of 
the naive perturbation expansion.

First, 
to collect secular terms in \fr{eqn:A2-sol-i} perturbatively one defines 
the following polynomials
\beqa
A_{\chN}^{(pol,-)}(t)&:=&A_{\chN}^{(0,-)}-i\epsilon^2 t\,
\Omega_{\chN}^{(ren,\,2)} \,A_{\chN}^{(0,-)}, 
\label{eqn:rgt-i}\\
A_{\chN}^{(pol,+)}(t)&:=&A_{\chN}^{(0,+)}+i\epsilon^2 t\, 
\Omega_{\chN}^{(ren,\,2)}\,A_{\chN}^{(0,+)},
\label{eqn:rgt-i+}
\eeqa
where 
$$
\Omega_{\chN}^{(ren,\,2)}:=-
\frac{1}{L\Omega_{\chN}}\left(\frac{c}{\NN{N}}\right)^2
\bigg[
\sum_{\eta'}\sum_{n'=n\pm 1}\sum_{q'(n')}
F_{\chN,\chN'}^{(1)}\frac{2}{L}\left(\frac{c}{\NN{N'}}\right)^2
\frac{F_{\chN',\chN}^{(1)}}{\Omega_{\chN'}^2-\Omega_{\chN}^2}
+F_{\chN,\chN}^{(2)}\bigg].\quad \in\mathbb{R}
$$
From these definitions, the naive solution can approximately be written as 
\beqa
A_{\chN}(\epsilon,t)&=&A_{\chN}^{(0)}(t)+\epsilon A_{\chN}^{(1)}(t)
+\epsilon^2 A_{\chN}^{(2)}(t)+{\cal O}(\epsilon^3)\non\\
&=&A_{\chN}^{(pol,-)}(t)\, e^{-i\Omega_{\chN}t}
+A_{\chN}^{(pol,+)}(t)\, e^{i\Omega_{\chN}t}
+{\cal O}(\epsilon^3)+(\,\mbox{NS}\,).
\label{eqn:relation-rg}
\eeqa

Second, one derives the equations which 
\fr{eqn:rgt-i} and \fr{eqn:rgt-i+} should
perturbatively satisfy. From the definitions one has  
$$
\frac{A_{\chN}^{(pol,\mp)}(t+\tau)-A_{\chN}^{(pol,\mp)}(t)}{\tau}=
\mp i\epsilon^2 \Omega_{\chN}^{(ren,\,2)} A_{\chN}^{(0,\mp)}, 
$$
where $\tau$ is a real constant. The right hand sides contain 
$A_{\chN}^{(0,\mp)}$. 
To obtain the closed equations 
in terms of $A_{\chN}^{(pol,\mp)}(t)$, not $A_{\chN}^{(0,\mp)}$, 
one substitutes 
$A_{\chN}^{(0,\mp)}=A_{\chN}^{(pol,\mp)}(t)+{\cal O}(\epsilon^2)$ which are
the inverse of the
definitions \fr{eqn:rgt-i} and \fr{eqn:rgt-i+}, then 
$$
\frac{A_{\chN}^{(pol,\mp)}(t+\tau)-A_{\chN}^{(pol,\mp)}(t)}{\tau}=
\mp i\epsilon^2 \Omega_{\chN}^{(ren,\,2)} A_{\chN}^{(pol,\mp)}(t) +{\cal O}(\epsilon^4).
$$

Taking the limit, $\tau\to 0$, for the both sides one has 
the renormalization equations 
\beqa
\frac{d }{dt}A_{\chN}^{(ren,\mp)}&=&
\mp i\epsilon^2 \Omega_{\chN}^{(ren,\,2)}A_{\chN}^{(ren,\mp)}
\label{eqn:rge}
\eeqa
where $A_{\chN}^{(ren,\mp)}(t)$ approximate $A_{\chN}^{(pol,\mp)}(t)$, and 
the solutions to \fr{eqn:rge} contain higher order terms in
$\epsilon$. The explicit forms of the 
solutions are     
\beq
A_{\chN}^{(ren,\mp)}(t)
=A_{\chN}^{(ren,\mp)}(0)e^{\mp i\epsilon^2\Omega_{\chN}^{(ren,\,2)}t}.
\label{eqn:rge-sol}
\eeq

Finally one obtains an approximate solution for
$A_{\chN}(\epsilon,t)$ that does not contain the secular terms.
Using \fr{eqn:relation-rg} and  \fr{eqn:rge-sol} one has 
$$
A_{\chN}(\epsilon, t)\approx A_{\chN}^{(ren,-)}(0)\,
e^{-i\Omega_{\chN}^{(ren)}t}
+A_{\chN}^{(ren,+)}(0)\,
e^{i\Omega_{\chN}^{(ren)}t}.
$$
The frequency-shift due to the curved boundary is obtained as 
$$
\Omega_{\chN}^{(ren)}:=\Omega_{\chN}+\epsilon^2\Omega_{\chN}^{(ren,\,2)}
+{\cal O}(\epsilon^3).
$$
\subsection{case (ii)}
We give the renormalization equations for the case (ii). 
The procedure to obtain these equations is same as 
in \S\ref{sec:renorm-case-i}.

To collect the secular terms one defines
\beqa
A_{\chN}^{(pol,\mp)}(t)&:=&A_{\chN}^{(0,\mp)}+\epsilon^2
\frac{2}{L}\left(\frac{c}{\NN{N}}\right)^2
\sum_{\eta'}\sum_{n'=n\pm 1}\sum_{q'(n')}
F_{\chN,\chN'}^{(1)}\frac{2}{L}
\left(\frac{c}{\NN{N'}}\right)^2
\non\\
&&\qquad\quad  
\times\,
\sum_{\eta''}^{(\res)}\sum_{n''=n'\pm 1}^{(\res)}\sum_{q''(n'')}^{(\res)}
\frac{t}{\mp 2i\Omega_{\chN}}
\frac{F_{\chN',\chN''}^{(1)}}{\Omega_{\chN'}^2-\Omega_{\chN}^2}
A_{\chN}^{(0,\mp)}\non\\
&&+\epsilon^2\frac{2}{L}\left(\frac{c}{\NN{N}}\right)^2
\sum_{\eta''}^{(\res)}\sum_{n''=n'\pm 1}^{(\res)}\sum_{q''(n'')}^{(\res)}
\frac{t }{\mp2i\Omega_{\chN}}
F_{\chN,\chN''}^{(2)}
A_{\chN''}^{(0,\mp)}.
\non
\eeqa
From these equations one can obtain the renormalization equations 
\beqa
\frac{dA_{\chN}^{(ren,\mp)}}{dt}&=&\epsilon^2
\frac{2}{L}\left(\frac{c}{\NN{N}}\right)^2
\sum_{\eta'}\sum_{n'=n\pm 1}\sum_{q'(n')}
F_{\chN,\chN'}^{(1)}\frac{2}{L}
\left(\frac{c}{\NN{N'}}\right)^2\non\\
&&\times
\sum_{\eta''}^{(\res)}\sum_{n''=n'\pm 1}^{(\res)}\sum_{q''(n'')}^{(\res)}
\frac{1}{\mp 2i\Omega_{\chN}}
\frac{F_{\chN',\chN''}^{(1)}}{\Omega_{\chN'}^2-\Omega_{\chN}^2}
A_{\chN}^{(ren,\mp)}\non\\
&&+\epsilon^2\frac{2}{L}\left(\frac{c}{\NN{N}}\right)^2
\sum_{\eta''}^{(\res)}\sum_{n''=n'\pm 1}^{(\res)}\sum_{q''(n'')}^{(\res)}
\frac{1}{\mp 2i\Omega_{\chN}}
F_{\chN,\chN''}^{(2)}
A_{\chN''}^{(ren,\mp)}.
\label{eqn:rge-ii}
\eeqa
Unlike the case (i), the right hand side of \fr{eqn:rge-ii}
consists of some different mode amplitudes, 
then the frequency-shift cannot directly be obtained from this
form of the renormalization equations.  
One can rewrite  \fr{eqn:rge-ii} as  
\beq
\frac{d}{dt}A_{\chN}^{(ren,\mp)}=\pm i \epsilon^2 \sum_{\chN'}
M_{\chN,\chN'}^{(ii)}
A_{\chN'}^{(ren,\mp)},
\label{eqn:M}
\eeq
where $\{M_{\chN,\chN'}^{(ii)}\}$ is a real matrix.
\subsection{case (iii)}
Finally, defining
\beqa
&&A_{\chN}^{(pol,\mp)}(t):=A_{\chN}^{(0,\mp)}+\epsilon
\frac{2}{L}\left(\frac{c}{\NN{N}}\right)^2
\sum_{\eta'}^{(\res)} \sum_{n'=n\pm 1}^{(\res)} \sum_{q'(n')}^{(\res)}
\frac{F_{\chN,\chN'}^{(1)}}{\mp 2i\Omega_{\chN} }
tA_{\chN'}^{(0,\mp)}
\non\\
&&+\epsilon^2 
\frac{2}{L}\left(\frac{c}{\NN{N}}\right)^2
\sum_{\eta'}^{(\res)}\sum_{n'=n\pm 1}^{(\res)}\sum_{q'(n')}^{(\res)}
F_{\chN,\chN'}^{(1)}\frac{2}{L}
\left(\frac{c}{\NN{N'}}\right)^2
\non\\
&&\qquad
\times\,
\sum_{\eta''}^{(\res)}\sum_{n''=n'\pm 1}^{(\res)}\sum_{q''(n'')}^{(\res)}
\frac{F_{\chN',\chN''}^{(1)}}{-2i\Omega_{\chN'}}
\left(\frac{\pm it^2}{4\Omega_{\chN}}+\frac{t}{4\Omega_{\chN}^2}\right)
\, A_{\chN'}^{(0,\mp)}\non\\
&&\quad
+\epsilon^2\frac{2}{L}\left(\frac{c}{\NN{N}}\right)^2
\sum_{\eta''}^{(\res)}\sum_{n''=n'\pm 1}^{(\res)}\sum_{q''(n'')}^{(\res)}
\frac{t}{\mp 2i\Omega_{\chN}}
F_{\chN,\chN''}^{(2)}
A_{\chN''}^{(0,\mp)},
\label{eqn:renorm-def-iv}
\eeqa
one has 
\beqa
&&\frac{dA_{\chN}^{(ren,\mp)}}{dt}=\epsilon
\frac{2}{L}\left(\frac{c}{\NN{N}}\right)^2
\sum_{\eta'}^{(\res)} \sum_{n'=n\pm 1}^{(\res)} \sum_{q'(n')}^{(\res)}
\frac{F_{\chN,\chN'}^{(1)}}{\mp 2i\Omega_{\chN} }
A_{\chN'}^{(ren,\mp)}
\non\\
&&+\epsilon^2 
\frac{2}{L}\left(\frac{c}{\NN{N}}\right)^2
\sum_{\eta'}^{(\res)}\sum_{n'=n\pm 1}^{(\res)}\sum_{q'(n')}^{(\res)}
F_{\chN,\chN'}^{(1)}\frac{2}{L}
\left(\frac{c}{\NN{N'}}\right)^2
\sum_{\eta''}^{(\res)}\sum_{n''=n'\pm 1}^{(\res)}\sum_{q''(n'')}^{(\res)}
\frac{F_{\chN',\chN''}^{(1)}}{\mp 2i\Omega_{\chN'}}
\frac{1}{4\Omega_{\chN}^2}\, A_{\chN'}^{(ren,\mp)}\non\\
&&+\epsilon^2\frac{2}{L}\left(\frac{c}{\NN{N}}\right)^2
\sum_{\eta''}^{(\res)}\sum_{n''=n'\pm 1}^{(\res)}\sum_{q''(n'')}^{(\res)}
\frac{1}{\mp 2i\Omega_{\chN}}
F_{\chN,\chN''}^{(2)}
A_{\chN''}^{(ren,\mp)}.
\label{eqn:rge-iv}
\eeqa
Here in the course of deriving the renormalization equations  
we have substituted 
$$
A_{\chN}^{(0,\mp)}=A_{\chN}^{(pol,\mp)}(t)-\epsilon t
\frac{2}{L}\left(\frac{c}{\NN{N}}\right)^2
\sum_{\eta'}^{(\res)} \sum_{n'=n\pm 1}^{(\res)} \sum_{q'(n')}^{(\res)}
\frac{F_{\chN,\chN'}^{(1)}}{\mp 2i\Omega_{\chN} }
A_{\chN'}^{(pol)}(t)+{\cal O}(\epsilon^2),
$$
which are from the definitions of $A_{\chN}^{(pol,\mp)}(t)$,  
\fr{eqn:renorm-def-iv}. 
One can rewrite \fr{eqn:rge-iv} in the matrix form as 
$$
\frac{d}{dt}A_{\chN}^{(ren,\mp)}=\pm i\epsilon
\sum_{\chN'}M_{\chN,\chN'}^{(iii)}A^{(ren,\mp)},
$$
where $\{M_{\chN,\chN'}^{(iii)}\}$ is a real matrix, as same as 
the case of (ii).

\section{Discussion and Conclusions}

In this paper we have explored physical effects due to 
non-trivial spatial boundary conditions   
for the system of a linear wave equation. As a simple example, 
we have studied the system 
with Dirichlet type boundary conditions in a prescribed weakly curved pipe. 
The perturbation scheme was based on the non-perturbed eigenvalue problems 
of the Laplacians and gave us the set of ordinary differential equations.
Then it has been observed that self-mode couplings occur
and then secular terms appear at the second order analysis, in
addition to the secular terms due to resonances between different modes.
The resonance conditions have been derived at each order in the 
naive perturbative analysis and it 
turns out that these resonance conditions do not 
contain $\kappa(z)$.
Using the renormalization method we have derived amplitude equations.
In the case where self-mode coupling only occurs we 
have obtained an analytical 
expression for the frequency-shift due to the curved boundary. 
In the case where 
there are resonances between different modes, 
we have derived the amplitude equations in a matrix form as 
renormalization equations. 
The amplitude equation with higher order correction terms 
can be obtained by applying our 
procedure straightforwardly in each case. 
Beyond this study, our methodology can be applied to systems with 
other boundary conditions, for example, the prescribed curve is not planer and 
the cross section is rectangular. In the case where the cross section
is rectangular, Bessel functions as the eigenfunctions for 
the two-dimensional Laplacian will be replaced with sinusoidal functions.
Furthermore, if the given system has a weakly nonlinear term,
there will be nonlinear resonance terms in the naive perturbation
expansion in addition to those associated with curved boundary.
%
Thus it is obvious from our procedure to obtain the amplitude equations 
that the corresponding amplitude equations will be nonlinear. 
We believe that our present work and these extensions that follow 
from this work can help 
elucidate the behavior of systems with non-trivial spacial boundary conditions.

%


\section{Appendix}
\label{appendix}
In this appendix we study the following problems
\beq
\ddot{A}+\Omega^2 A=
 ( \alpha_1^{(-)}t + \alpha_0^{(-)})\,e^{-i\Omega t}
+( \alpha_1^{(+)}t + \alpha_0^{(+)})\,e^{i\Omega t},
\label{EqnA-reso}
\eeq
and 
\beq
\ddot{A}+\Omega^2 A=
 ( \beta_1^{(-)}t + \beta_0^{(-)})\,e^{-i\Omega' t}
+( \beta_1^{(+)}t + \beta_0^{(+)})\,e^{i\Omega' t},
\label{EqnA-nonreso}
\eeq
Here $\dot{}$ denotes the differentiation with respect to $t$, 
$\alpha_j^{(\pm)},\beta_j^{(\pm)}\in\mathbb{C}$ and 
$\Omega,\Omega'(\neq\pm\Omega)\in\mathbb{R}$
are given constants.

The solution to \fr{EqnA-reso} is obtained as 
$$
A(t)=\left\{\frac{\alpha_1^{(-)}}{-4i\Omega}t^2+
\left(\frac{\alpha_0^{(-)}}{-2i\Omega}+\frac{\alpha_1^{(0)}}{4\Omega^2}
\right)t\right\}\,e^{-i\Omega t}
+
\left\{\frac{\alpha_1^{(+)}}{4i\Omega}t^2+
\left(\frac{\alpha_0^{(+)}}{2i\Omega}+\frac{\alpha_1^{(0)}}{4\Omega^2}
\right)t\right\}\,e^{i\Omega t}.
$$
 Similarly, the solution to \fr{EqnA-nonreso} is obtained as 
\beqa
A(t)&=&\left\{\frac{\beta_1^{(-)}}{\Omega^2-\Omega^{\prime 2}}t+
\frac{\beta_0^{(-)}}{\Omega^2-\Omega^{\prime 2}}
+2i\Omega'\frac{\beta_1^{(-)}}{(\Omega^2-\Omega^{\prime 2})^2}
\right\}\,e^{-i\Omega' t}\non\\
&&+
\left\{\frac{\beta_1^{(+)}}{\Omega^2-\Omega^{\prime 2}}t+
\frac{\beta_0^{(-)}}{\Omega^2-\Omega^{\prime 2}}
-2i\Omega'\frac{\beta_1^{(+)}}{(\Omega^2-\Omega^{\prime 2})^2}
\right\}\,e^{i\Omega' t}
.\non
\eeqa



\vspace{5mm}

\end{document}